\documentclass[a4paper,11pt]{ijamas}

\usepackage{times}
\usepackage{graphicx}

\newcommand{\R}{{\rm I}\!{\rm R}} % the set of real numbers

\begin{document}

%%%%%%%%%%%%%%%%%%%%%%%%%%%%%%%%%%%%%%%%%%%%%%%%%%%%%%

\title{Shape of Traveling Densities with \\ Extremum Statistical Complexity}

\author{\textbf{Ricardo L\'opez-Ruiz$^1$ and Jaime Sa\~nudo$^2$}}

\date{$^1$Department of Computer Science and BIFI\\
Universidad de Zaragoza\\
Campus San Francisco, E-50009 Zaragoza, Spain\\
rilopez@unizar.es\\[0.3cm]
$^2$Department of Physics and BIFI\\
Universidad de Extremadura\\
Avda. de Elvas, E-06071 Badajoz, Spain\\
jsr@unex.es}

\maketitle

%%%%%%%%%%%%%%%%%%%%%%%%%%%%%%%%%%%%%%%%%%%%%%%%%%%%%%

\begin{abstract}
\noindent \emph{In this paper, we analyze the behavior of statistical complexity in several systems 
where two identical densities that travel in opposite direction cross each other.
Besides the crossing between two Gaussian, rectangular and triangular densities studied
in a previous work, we also investigate in detail the crossing between two exponential 
and two gamma distributions. For all these cases, 
the shape of the total density presenting an extreme value in complexity is found.}

\medskip


\medskip


\end{abstract}

%%%%%%%%%%%%%%%%%%%%%%%%%%%%%%%%%%%%%%%%%%%%%%%%%%%%%%%

\section{Introduction}
\label{S1}

The behavior of the statistical complexity
in time-dependent systems has not been broadly investigated. 
A previous work in this direction is presented in
\cite{calbet2001} where a gas decaying toward the 
asymptotic equilibrium state is studied.
It was found that this system goes towards equilibrium by 
approaching the {\it maximum complexity path}, which is the trajectory
in distribution space formed by the distributions with the maximal
complexity. Then, from a physical point of view, it can have some interest
to study the extremal behavior of statistical magnitudes in time dependent systems.

In this work, we start by studying  the statistical 
complexity $C$ in a simplified time-dependent system $\rho(x,t)$ composed of 
two one-dimensional (variable $x$) identical densities that travel 
in opposite directions with the same velocity $v$, one of them, 
$\rho_+(x,t)$, going to the right and the other one, $\rho_-(x,t)$ 
going to the left. That is  
\begin{equation}
\rho(x,t)={1\over 2}\,\rho_+(x,t)+{1\over 2}\,\rho_-(x,t),
\label{eq-rho}
\end{equation}
with the normalization condition $\int_{\R}\rho_{\pm}(x,t)dx=1$
that implies the normalization of $\rho(x,t)$, and the initial condition 
$\rho_+(x,0)=\rho_-(x,0)$.
In the next section, we recall the analysis of $C$ done for two Gaussian, 
rectangular and triangular traveling densities in \cite{tunez2010}.
Also, the extension of this study for two exponential and two gamma distributions
is performed. Specifically, the shape of $\rho(x,t)$ presenting
the maximum and minimum $C$ is explicitly shown for all these cases. The final section
includes our conclusions.

%%%%%%%%%%%%%%%%%%%%%%%%%%%%%%%%%%%%%%%%%%%%%%%%%%%%%%%

\section{Complexity in Traveling Densities}
\label{S2} \vspace{-4pt}

Let us start by recalling the definition of the statistical complexity $C$ \cite{lopez1995},
the so-called $LMC$ complexity, that is defined as
\begin{equation}
C = H\cdot D\;,
\end{equation}
where $H$ represents the information content of the system and $D$ gives an idea
of how much concentrated is its spatial distribution. 
For our purpose, we take a version used in \cite{lopez2002}
as quantifier of $H$. This is the simple exponential Shannon entropy \cite{dembo1991},
that takes the form, 
\begin{equation}
H = e^{S}\;,
\end{equation}
where $S$ is the Shannon information entropy \cite{shannon1948},
\begin{equation}
S = -\int \rho(x)\;\log \rho(x)\; dx \;,
\label{eq1}
\end{equation}
with $x$ representing the continuum of the system states and $\rho(x)$
the probability density associated to all those states.
We keep for the disequilibrium the form originally introduced in 
\cite{lopez1995,lopez2002}, that is,
\begin{equation}
D = \int \rho^2(x)\; dx\;.
\label{eq2} 
\end{equation}
Now we proceed to calculate $C$ for the system above mentioned (\ref{eq-rho})
in the Gaussian, exponential, rectangular, triangular and gamma cases. 

\subsection{Gaussian traveling densities}

Here the two one-dimensional traveling densities that compose system (\ref{eq-rho}) 
take the form: 
\begin{equation}
\rho_{\pm}(x,t)={1\over \sigma\sqrt{2\pi}}\,\exp\left\{{-{(x\mp vt)^2}\over 2\sigma^2}\right\},
\label{eq-gauss}
\end{equation}
where $\sigma$ is the variance of the density distribution.

\begin{figure}[]  
\centerline{\includegraphics[width=9cm]{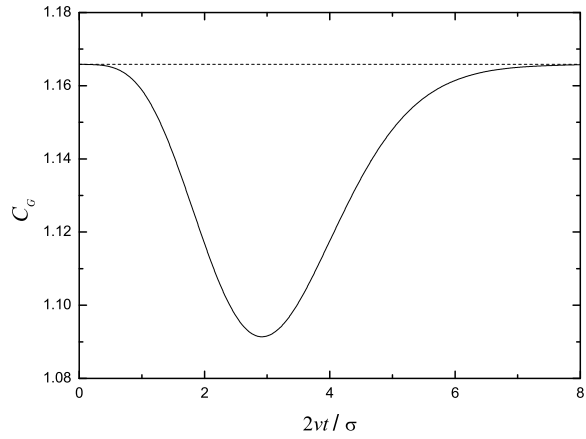}}     
\caption{\small Statistical complexity, $C_G$, vs. the adimensional separation, $2vt/\sigma$, 
between the two traveling Gaussian densities defined in Eq. (\ref{eq-gauss}).
The minimum of $C_G$ is reached when $2vt/\sigma=2.91$.
The dashed line indicates the value of complexity for the normalized Gaussian distribution.}  
\label{fig1}  
\end{figure}

\begin{figure}[]  
\centerline{\includegraphics[width=9cm]{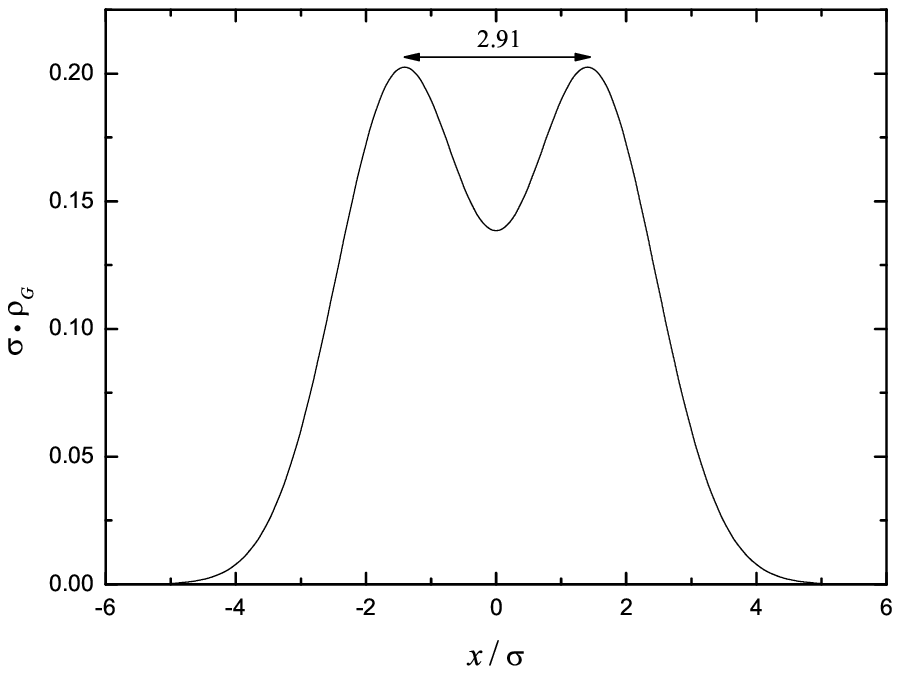}}     
\caption{\small Shape of the density (\ref{eq-rho}) in adimensional units that presents the 
minimum statistical complexity when the two traveling Gaussian densities defined in (\ref{eq-gauss}) 
are crossing. Notice that the value of the adimensional
separation between the centers of both Gaussian distributions must be $2.91$.}  
\label{fig2}  
\end{figure}

The behavior of complexity, $C_G$, as a function of the adimensional quantity $2vt/\sigma$ is given 
in Fig. \ref{fig1}. Let us observe that $C_G$ presents a minimum. The shape of system (\ref{eq-rho})
for this minimum complexity case is plotted in an adimensional scale in Fig. \ref{fig2}.

\begin{figure}[]  
\centerline{\includegraphics[width=9cm]{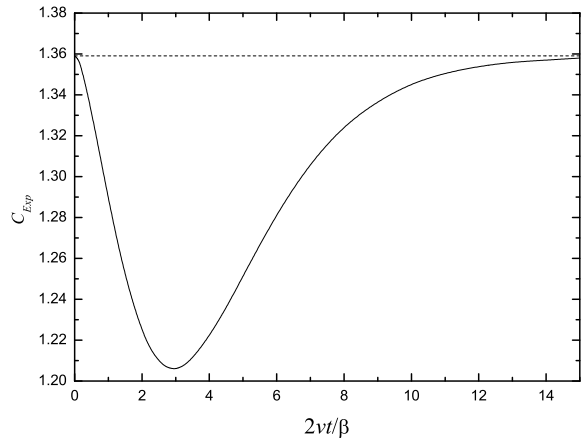}}     
\caption{\small Statistical complexity, $C_{Exp}$, vs. the adimensional separation, $2vt/\beta$, 
between the two traveling exponential densities defined in Eq. (\ref{eq-exp}).
The minimum of $C_{Exp}$ is reached when $2vt/\beta=2.94$.
The dashed line indicates the value of complexity for the normalized exponential distribution.}  
\label{fig8}  
\end{figure}  

\begin{figure}[]  
\centerline{\includegraphics[width=9cm]{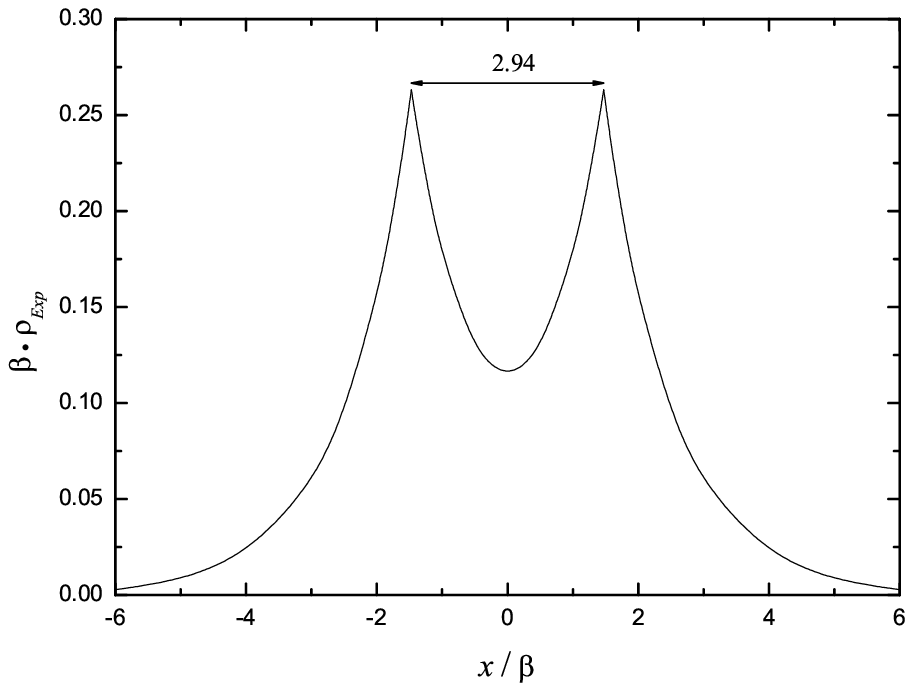}}     
\caption{\small Shape of the density (\ref{eq-rho}) in adimensional units that presents the 
minimum statistical complexity when the two traveling exponential densities defined in (\ref{eq-exp}) 
are crossing. Notice that the value of the adimensional
separation between the centers of both exponential distributions must be $2.94$.}  
\label{fig9}  
\end{figure}

\subsection{Exponential traveling densities}

A similar case to the Gaussian former one is the crossing between two exponential distributions.
Here the two one-dimensional traveling densities that compose system (\ref{eq-rho}) 
take the form: 
\begin{equation}
\rho_{\pm}(x,t)=
{1\over 2\beta}\exp\left\{-|x\mp vt|/\beta\right\} \,,
\label{eq-exp}
\end{equation}
where $\beta$ is the width of the density distribution.

The formula for $C_{Exp}$ can be analytically obtained.
The behavior of $C_{Exp}$ as a function of the adimensional quantity 
$2vt/\beta$ is given in Fig. \ref{fig8}. 
Let us observe that $C_{Exp}$ presents a minimum. The shape of system (\ref{eq-rho})
for this minimum complexity case is plotted in an adimensional scale in Fig. \ref{fig9}.

\subsection{Rectangular traveling densities}

Now the two one-dimensional traveling densities that compose system (\ref{eq-rho}) 
take the form: 
\begin{equation}
\rho_{\pm}(x,t)=\left\{
\begin{array}{llc}
1/\delta & \mbox{if} & -\delta/2\leq x\mp vt\leq\delta/2\,, \\
& \\
0 & \mbox{if} & |x\mp vt|>\delta/2\,.
\end{array}
\right.
\label{eq-rectang}
\end{equation}
where $\delta$ is the width of each distribution.

For this case, the complexity, $C_R$, can be analytically obtained. Its expression is:
\begin{equation}
C_R(t)=\left\{
\begin{array}{llc}
2^{2vt/\delta}\left(1-{vt\over \delta}\right)& \mbox{if} & 0\leq 2vt \leq\delta\,, \\
& \\
1 & \mbox{if} & 2vt >\delta\,,
\end{array}
\right.
\label{eq-C-rect}
\end{equation}

\begin{figure}[]  
\centerline{\includegraphics[width=9cm]{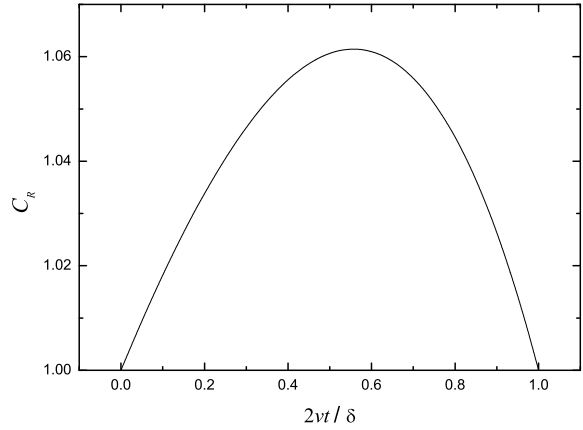}}     
\caption{\small Statistical complexity, $C_R$, vs. the adimensional separation, $2vt/\delta$, 
between the two traveling rectangular densities defined in Eq. (\ref{eq-rectang}).
The maximum of $C_R$ is reached when $2vt/\delta=0.557$.
Observe that the normalized rectangular distribution has $C_R=1$.}  
\label{fig3}  
\end{figure}  

\begin{figure}[]  
\centerline{\includegraphics[width=9cm]{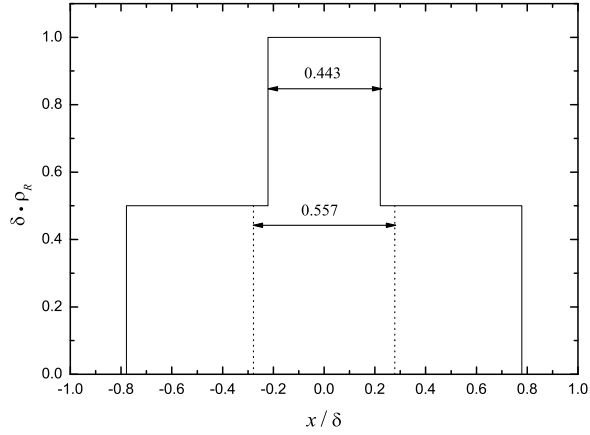}}     
\caption{\small Shape of the density (\ref{eq-rho}) in adimensional units that presents 
the maximum statistical complexity when the two traveling rectangular densities defined 
in (\ref{eq-rectang}) are crossing. Notice that the value of the adimensional separation 
between the centers of both rectangular distributions must 
be $0.557$. Then, the width of the overlapping between both distributions is 0.443.}  
\label{fig4}  
\end{figure}  

The behavior of $C_R$ as a function of the adimensional quantity $2vt/\delta$ is given 
in Fig. \ref{fig3}. Let us observe that $C_R$ presents a maximum. The shape of system (\ref{eq-rho})
for this maximun complexity case is plotted in an adimensional scale in Fig. \ref{fig4}.

\subsection{Triangular traveling densities}

The two one-dimensional traveling densities that compose system (\ref{eq-rho}) 
take the form in this case: 
\begin{equation}
\rho_{\pm}(x,t)=\left\{
\begin{array}{llc}
{(x\mp vt)\over\epsilon^2} + {1\over\epsilon} & \mbox{if} & -\epsilon\leq x\mp vt\leq 0 \,, \\
& & \\
{-(x\mp vt)\over\epsilon^2} + {1\over\epsilon} & \mbox{if} & 0 < x\mp vt\leq\epsilon \,, \\
& & \\
0 & \mbox{if} & |x\mp vt|>\epsilon \,,
\end{array}
\right.
\label{eq-triang}
\end{equation}
where $\epsilon$ is the width of each distribution (isosceles triangle whose base length is $2\epsilon$ ).

\begin{figure}[] 
\centerline{\includegraphics[width=9cm]{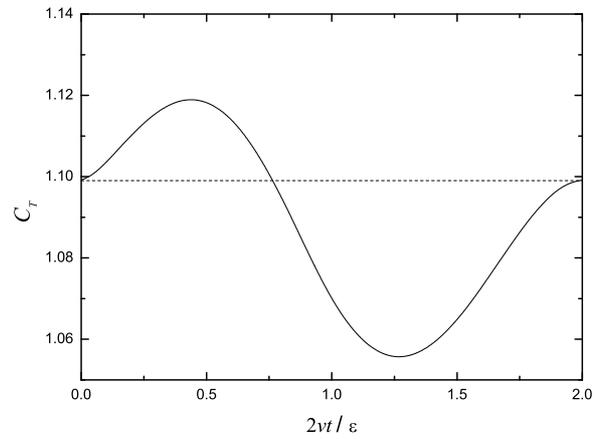}}     
\caption{\small Statistical complexity, $C_T$, vs. the adimensional separation, $2vt/\epsilon$, 
between the two traveling triangular densities given in Eq. (\ref{eq-triang}).
The maximum and minimum of $C_T$ are reached when $2vt/\epsilon$ takes the values $0.44$ and $1.27$,
respectively. The dashed line indicates the value of complexity for the normalized triangular distribution.}  
\label{fig5}  
\end{figure}

\begin{figure}[]  
\centerline{\includegraphics[width=9cm]{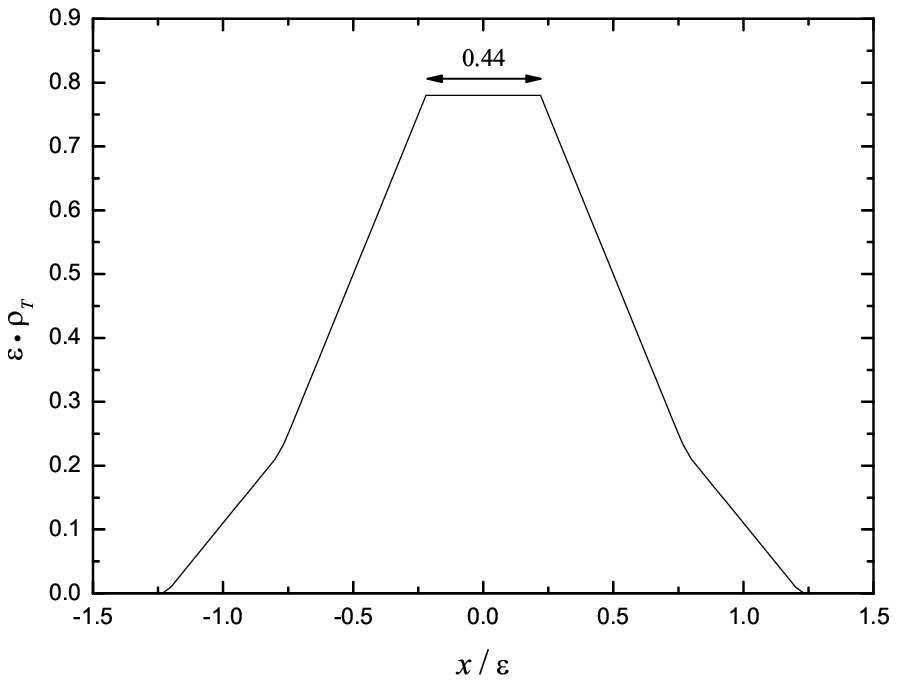}}     
\caption{\small Shape of the density (\ref{eq-rho}) in adimensional units that presents 
the maximum statistical complexity
when the two traveling triangular densities defined in (\ref{eq-triang}) are crossing. 
Notice that the value of the adimensional separation between the centers of both triangular 
distributions must be $0.44$.}  
\label{fig6}  
\end{figure}

\begin{figure}[]  
\centerline{\includegraphics[width=9cm]{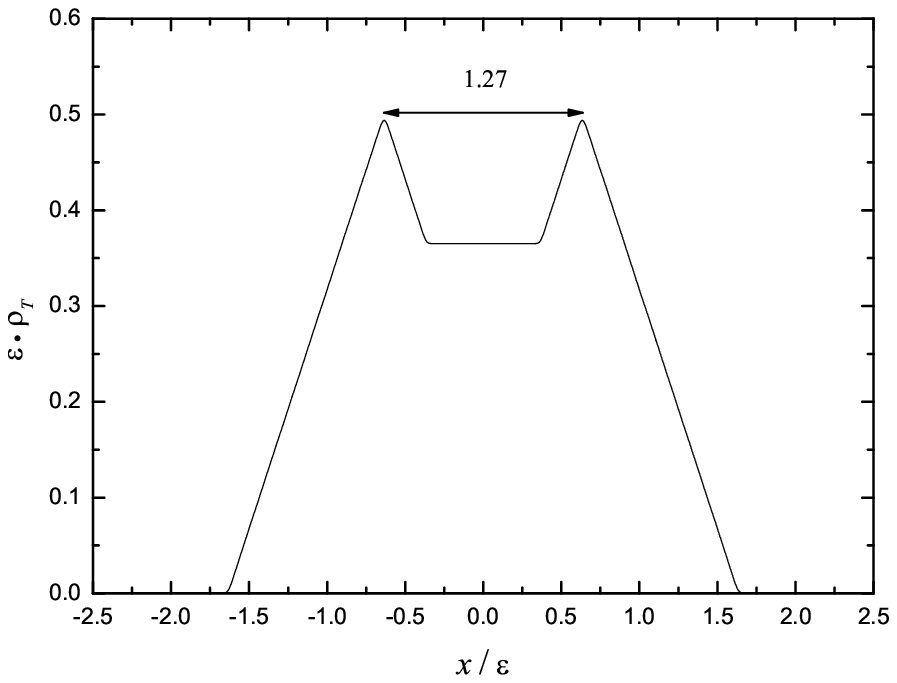}}     
\caption{\small Shape of the density (\ref{eq-rho}) in adimensional units that presents 
the minimum statistical complexity
when the two traveling triangular densities defined in (\ref{eq-triang}) are crossing. 
Notice that the value of the adimensional separation between the centers of both triangular 
distributions must be $1.27$.}  
\label{fig7}  
\end{figure}

The behavior of complexity, $C_T$, as a function of the adimensional quantity $2vt/\epsilon$ is given 
in Fig. \ref{fig5}. Let us observe that $C_T$ presents a maximum and a minimum. The shape of system (\ref{eq-rho})
for both cases, with maximum and minimum complexitiy, are plotted in an adimensional scale in 
Figs. \ref{fig6} and \ref{fig7}, respectively.

\subsection{Gamma traveling densities}

We study a final case given by the crossing between two gamma distributions.
Here the two one-dimensional traveling densities that compose system (\ref{eq-rho}) 
take the form: 
\begin{equation}
\rho_{\pm}(x,t)={1\over 2\theta^k\Gamma(k)}\,|x\mp vt|^{k-1}
\exp\left\{{-{|x\mp vt|}/\theta}\right\},
\label{eq-gamma}
\end{equation}
where $k$ is a real parameter such that $k\geq 1$ and $\theta$ is a scale parameter related 
to the width of the distribution for a given $k$. Observe that the exponential case
is recovered for $k=1$. Let us also remark that for a fixed $k$ the parameter $\theta$
modifies the shape of the distribution but the complexity $C$ is conserved.

\begin{figure}[]  
\centerline{\includegraphics[width=9cm]{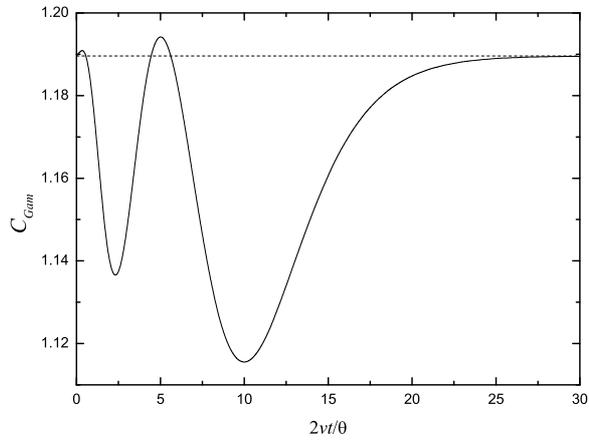}}     
\caption{\small Statistical complexity, $C_{Gam}$, vs. the adimensional separation, $2vt/\theta$, 
between the two traveling gamma densities given in Eq. (\ref{eq-gamma}) for $k=3$.
The local maxima of $C_{Gam}$ are reached when $2vt/\theta$ take the values $0.36$ and $5.01$,
and the local minima of $C_{Gam}$ are on the values $2.32$ and $9.99$. The dashed line indicates 
the value of complexity for the normalized gamma distribution for $k=3$.}  
\label{fig10}  
\end{figure}

\begin{figure}[]  
\centerline{\includegraphics[width=9cm]{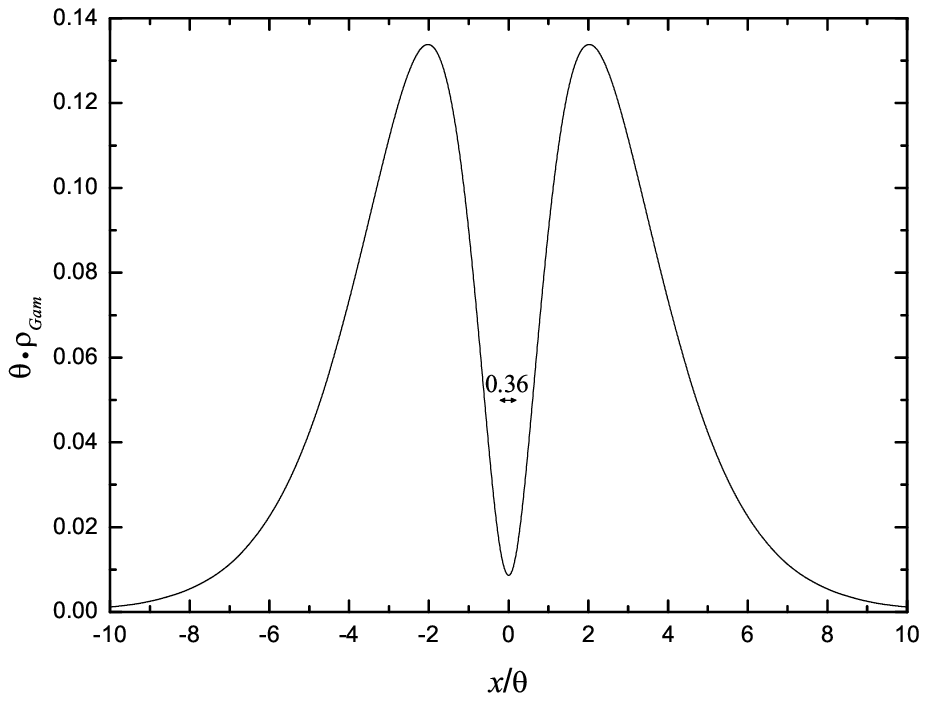}}     
\caption{\small Shape of the density (\ref{eq-rho}) in adimensional units that presents 
a local maximum statistical complexity
when the two traveling gamma densities defined in (\ref{eq-gamma}) are crossing. 
Notice that the value of the adimensional separation between the centers of both gamma 
distributions must be $0.36$.}  
\label{fig11}  
\end{figure}

\begin{figure}[]  
\centerline{\includegraphics[width=9cm]{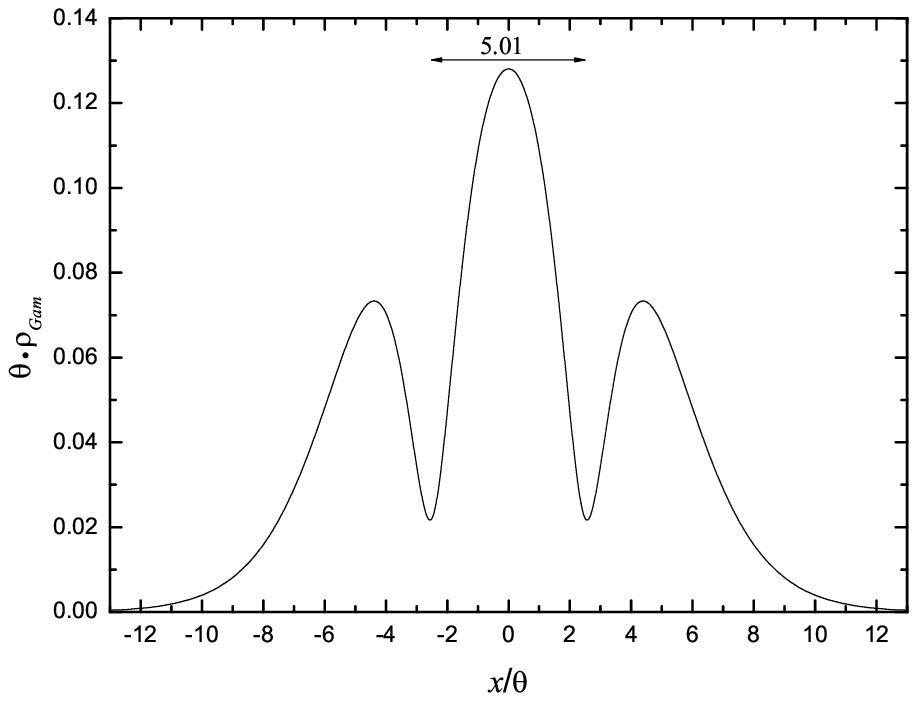}}     
\caption{\small Shape of the density (\ref{eq-rho}) in adimensional units that presents 
a local maximum statistical complexity when the two traveling gamma densities defined 
in (\ref{eq-gamma}) are crossing. Notice that the value of the adimensional separation 
between the centers of both gamma distributions must be $5.01$.}  
\label{fig12}  
\end{figure}

\begin{figure}[]  
\centerline{\includegraphics[width=9cm]{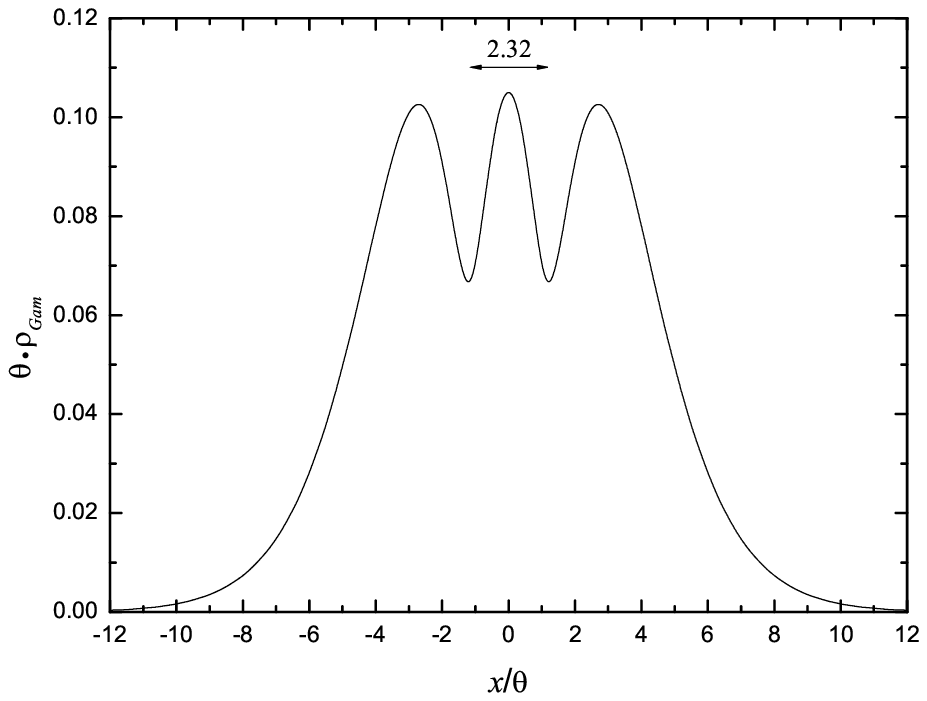}}     
\caption{\small Shape of the density (\ref{eq-rho}) in adimensional units that presents 
a local minimum statistical complexity when the two traveling gamma densities defined 
in (\ref{eq-gamma}) are crossing. Notice that the value of the adimensional separation 
between the centers of both gamma distributions must be $2.32$.}  
\label{fig13}  
\end{figure}

\begin{figure}[]  
\centerline{\includegraphics[width=9cm]{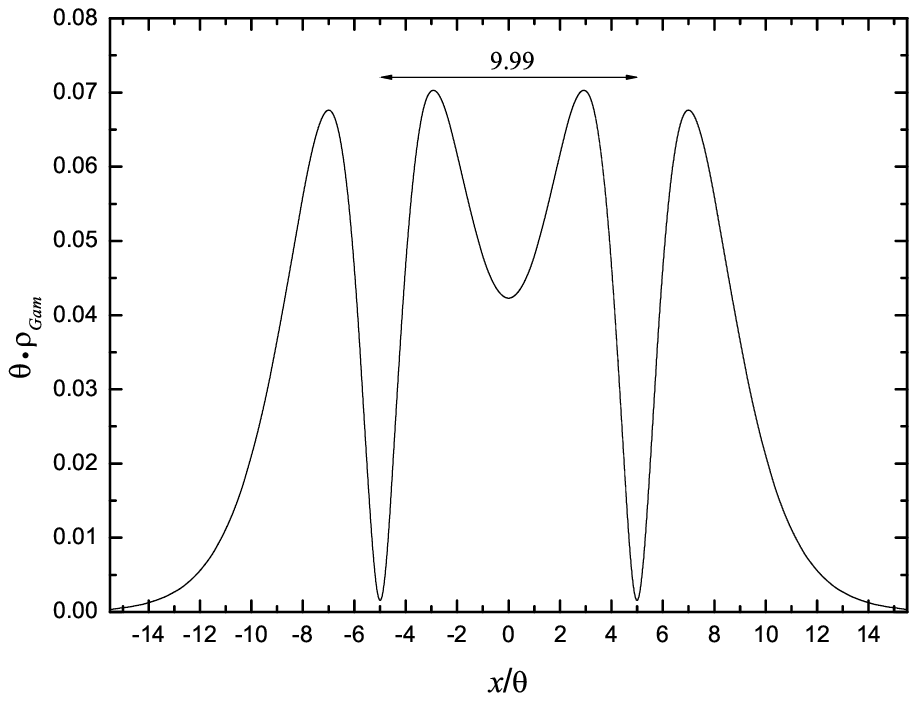}}     
\caption{\small Shape of the density (\ref{eq-rho}) in adimensional units that presents 
a local minimum statistical complexity when the two traveling gamma densities defined 
in (\ref{eq-gamma}) are crossing. Notice that the value of the adimensional separation 
between the centers of both gamma distributions must be $9.91$.}  
\label{fig14}  
\end{figure}

The behavior of complexity, $C_{Gam}$, as a function of the adimensional quantity $2vt/\theta$ is given 
in Fig. \ref{fig10} for $k=3$. The behavior for other values of $k$ seems to be similar.
Let us observe that $C_{Gam}$ presents two local maxima and two local minima. 
The shape of system (\ref{eq-rho}) for all these cases, with local extremal complexity, 
are plotted in an adimensional scale in Figs. \ref{fig11} and \ref{fig12}, for the local maxima,
and in Figs.  \ref{fig13} and \ref{fig14}, for the local minima.

\section{Conclusion}
\label{S3} 

In this work, we have studied the behavior of the statistical complexity
as a function of time when two traveling identical densities 
are crossing each other. Five cases have been analyzed: 
Gaussian, exponential, rectangular, triangular and gamma densities.
The Gaussian and exponential cases present a particular configuration with minimum complexity. 
The rectangular case displays a particular configuration with maximum complexity. 
The triangular case shows an intermediate behavior between the two former cases 
with two particular configurations, one of them with maximum complexity and the other one
with minimum complexity. The gamma case displays two configurations with local minima complexity
and other two configurations with local maxima complexity.
In general, all these configurations with extremal complexity cannot be analytically 
obtained and a careful computational study is required in order to determine them.

%%%%%%%%%%%%%%%%%%%%%%%%%%%%%%%%%%%%%%%%%%%%%%%%%%%%%%%

\section*{Acknowledgment}

This research was supported by the
spanish Grant with Ref. FIS2009-13364-C02-C01.
J.S. also thanks to the Consejer\'ia de Econom\'ia, Comercio e Innovaci\'on 
of the Junta de Extremadura (Spain) for financial support, Project Ref. GRU09011.

%%%%%%%%%%%%%%%%%%%%%%%%%%%%%%%%%%%%%%%%%%%%%%%%%%%%%%%

\bibliographystyle{dcu}
\bibliography{ijamas-C-time}

%%%%%%%%%%%%%%%%%%%%%%%%%%%%%%%%%%%%%%%%%%%%%%%%%%%%%%%

\end{document}